\documentclass[prl,twocolumn,showpacs,floatfix]{revtex4}

\usepackage{graphicx}
\usepackage{array}
\usepackage{amsmath,amsfonts,amssymb}
\usepackage[ansinew]{inputenc}
\usepackage{color}
\usepackage{calc}

\newcommand{\Figref}[1]{Fig.~\ref{#1}}

\begin{document}
\title{Electrically-driven vibronic spectroscopy with sub-molecular resolution}
\author
{Benjamin Doppagne$^1$, Michael C. Chong$^1$, Etienne Lorchat $^1$, St\'ephane Berciaud$^1$, Michelangelo Romeo$^1$, Herv\'e Bulou$^1$, Alex Boeglin$^1$,\\
 Fabrice Scheurer$^1$, Guillaume Schull$^{1\ast}$\\
\normalsize{$^1$ Universit\'e de Strasbourg, CNRS, IPCMS, UMR 7504, F-67000 Strasbourg, France,} \\
\altaffiliation{guillaume.schull@ipcms.unistra.fr} 
}

\begin{abstract}
A scanning tunneling microscope is used to generate the electroluminescence of phthalocyanine molecules deposited on NaCl/Ag(111). Photon spectra reveal an intense emission line at $\approx$ 1.9 eV that corresponds to the fluorescence of the molecules, and a series of weaker red-shifted lines. Based on a comparison with Raman spectra acquired on macroscopic molecular crystals, these spectroscopic features can be associated to the vibrational modes of the molecules and provide a detailed chemical fingerprint of the probed species. Maps of the vibronic features reveal sub-molecularly-resolved structures whose patterns are related to the symmetry of the probed vibrational modes.           
\end{abstract}

\date{\today}

\pacs{78.67.-n,78.60.Fi,68.37.Ef}

\maketitle

Near infrared \cite{Blanco2002}, Raman \cite{Gardiner1989} and low-temperature fluorescence \cite{Naumov2013} spectroscopies are powerful optical approaches to gather detailed chemical, structural or environmental information on organic systems. Extremely sensitive to vibrational modes, it is generally considered that these techniques provide an unambiguous fingerprint of the probed species. Conversely, the scanning tunneling microscope (STM) enables imaging organic or inorganic objects with atomic resolution but lacks chemical sensitivity. Pechenezhskiy \textit{et al.} recently developed an infrared STM able to detect the vibronic signature of monolayer assemblies of molecules \cite{Pechenezhskiy2013}. This method, however, could not yet reach single-molecule sensitivity. Nearly at the same time, Zhang \textit{et al.} made an important step towards combining the chemical sensitivity of Raman spectroscopy with the spatial resolution of the STM \cite{Zhang2013}. They performed tip-enhanced Raman spectroscopy (TERS) of molecules with sub-nanometric resolution using the tip of a STM as a plasmonic antenna to amplify a laser excitation.\\
Because they can be confined to atomic-scale pathways, we propose here to use electrons rather than photons as an excitation source of the vibronic signal. This approach is based on our recent observation of a large number of faint vibronic peaks in the low-temperature fluorescence spectra of a single-molecule excited by STM \cite{Chong2016,Chong2016a}. A comparison to calculated vibrational spectra led us to suggest that these faint peaks correspond to the different molecular vibration modes, in contrast to earlier STM-induced light emission (STM-LE) experiments where similar features pertained to the harmonic progression of a single mode \cite{Qiu2003,Wu2008,Chen2010,Zhu2013,Lee2014}. However, because of the low intensity of the signal and the lack of comparable spectroscopic data for this molecule in the literature, a definitive conclusion regarding the nature of the vibronic features could not be reached. Moreover, in this experiment, the molecular emitter is suspended between the tip and the sample of the STM junction, and the spatial resolution of the STM is lost.

With this work we demonstrate that spatial resolution and detailed vibronic spectroscopy can be combined in a single experiment without the need for an optical excitation. Following an experimental approach reported recently \cite{Zhang2016,Imada2016b,Imada2016}, we address the optoelectronic properties of zinc-phthalocyanine (ZnPc) molecules decoupled from a Ag(111) surface by a thin insulating layer of salt (NaCl). STM-LE spectra of these molecules reveal a sharp and intense emission line around 1.9 eV which corresponds to the fluorescence of the zinc phthalocyanine. Our spectra also show several weaker emission lines at lower energy. Extremely well reproduced by Raman spectra acquired on bulk ZnPc aggregates and fluorescence measurements on ZnPc trapped in frozen matrices \cite{Murray2011}, these emission lines are associated to the vibrational modes of the molecule and constitute an accurate spectroscopic fingerprint of the probed species. Sub-molecular spatial variations of the vibronic peak intensities were observed by scanning the molecule with the STM tip. In contrast to resonant TERS maps \cite{Zhang2013} that are insensitive to the probed vibronic mode \cite{Duan2015}, the patterns in our STM-LE vibronic maps reflect the symmetry of the considered modes, an effect that is interpreted in the framework of vibronic coupling theory\cite{Fisher1984}. \\

\noindent
The STM data were acquired with a low temperature (4.5\,K) Omicron setup operating in ultrahigh vacuum adapted to detect the light emitted at the tip-sample junction. The optical detection setup is composed of a grating spectrograph coupled to a cooled CCD camera and provides a spectral resolution of $\approx$ 1 nm \cite{Chong2016}. Tungsten STM-tips were introduced in the sample to cover them with silver and to tune their plasmonic response. The Ag(111) substrates were cleaned with successive sputtering and annealing cycles. Approximately 0.5 monolayer of NaCl was sublimed on Ag(111) kept at room temperature, forming square bi- and tri-layers. Eventually, zinc-phthalocyanine (ZnPc) molecules were evaporated on the cold ($\approx$ 5 K) NaCl/Ag(111) sample in the STM chamber. DFT calculations of single ZnPc were carried out at the B3LYP/6-13G level in the full D4h geometry to determine the vibrational modes and their symmetries \cite{Gaussian09}. The calculated vibrational frequencies were scaled by 0.9613 as recommended \cite{Liu2007}.\\

\noindent  
 
\begin{figure}
  \includegraphics[width=0.9\linewidth]{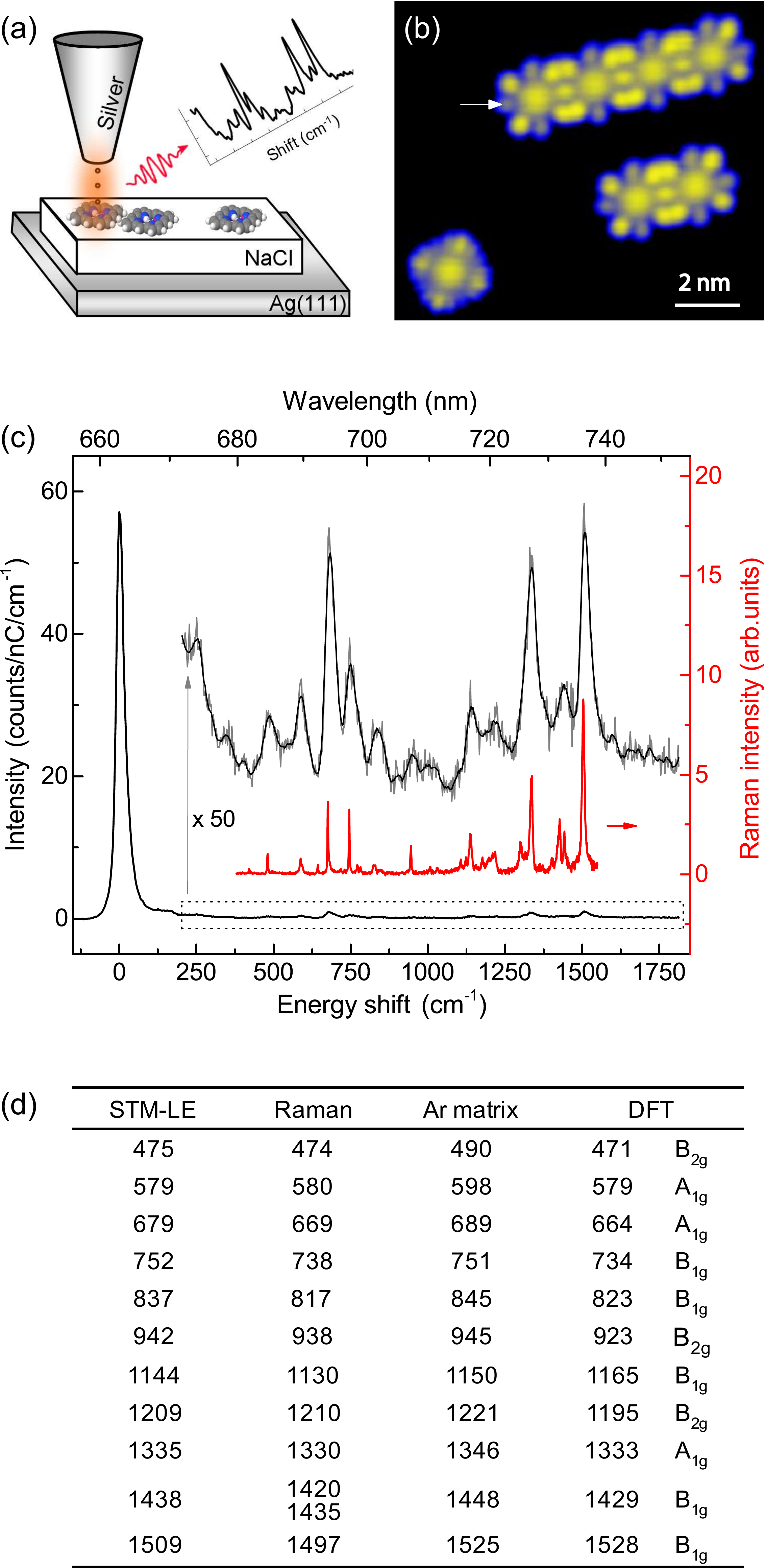}
  \caption{\label{fig1}(a) Sketch of the STM-induced emission experiment. (b) STM image ($I$ = 30 pA, $V$ = - 2.5 V) of a single ZnPc molecule (lower left), a dimer (lower right) and a linear tetramer (top right) of ZnPc molecules deposited on NaCl/Ag(111). (c) STM-induced light emission spectrum (black line) of the ZnPc linear tetramer ($I$ = 0.75 nA, $V$ = - 2.5 V, $t$ = 300 s, tip located at the white arrow in (b)). The right part of the spectrum is represented magnified by a factor of 50 and compared to a resonance Raman spectra acquired on a ZnPc crystal at room temperature and in air (excitation wavelength 632.8 nm). The raw (smoothed) data appear in grey (black) in the magnified spectrum. (d) Frequency shifts (in cm$^{-1}$) of the vibronic peaks in the STM-LE and Raman spectra displayed in (c), in the fluorescence spectra of ZnPc molecules trapped in frozen Argon matrices taken from Ref.\citenum{Murray2011} and assignment to vibrational modes of the ZnPc with their respective symmetry based on DFT calculations.} 
\end{figure}

Figure \ref{fig1}(a) is an illustration of the experiment where an STM tip is used to excite the luminescence of ZnPc molecules adsorbed on a 2ML-NaCl/Ag(111) substrate. Figure \ref{fig1}(b) is an STM image of a single molecule, a dimer and a tetramer of ZnPC molecules assembled on this surface. The molecular arrangements are formed by STM-tip manipulation following the procedure established in ref. [\citenum{Zhang2016}]. The optical spectrum in \Figref{fig1}(c) is obtained by locating the STM-tip on the extremity of the tetramer (see the arrow in \Figref{fig1}(b)) and applying a sample voltage of $V$ = - 2.5 V with a current setpoint $I$ = 0.75 nA.  The electroluminescence spectrum shows an intense emission line at 663 nm -- the 0-0 transition -- and several red-shifted peaks of lower intensity (magnified by a factor 50 in the top part of the panel). The main emission line is assigned to the emission of an excitonic state delocalized over several molecules \cite{Zhang2016}. The weaker emission lines (which were not reported in Ref.\citenum{Zhang2016}) are represented as a function of their energy shift from the 0-0 line chosen as origin of the abscissa. They are compared in \Figref{fig1}c to an experimental Raman spectrum of a bulk ZnPc crystal acquired for close-to resonance excitation. The number, the energies and the relative intensities of the different peaks are nearly identical in these two spectra. These data are also in excellent agreement with photoluminescence spectra acquired on ZnPc molecules isolated in cryogenic matrices \cite{Murray2011}. Based on a comparison with DFT calculations, the emission lines can be precisely assigned to distinct vibrational modes of the ZnPc molecule (\Figref{fig1}(d)). Our STM-induced light emission spectra therefore constitute a precise chemical fingerprint of the probed species.\\

\begin{figure}
  \includegraphics[width=1.00\linewidth]{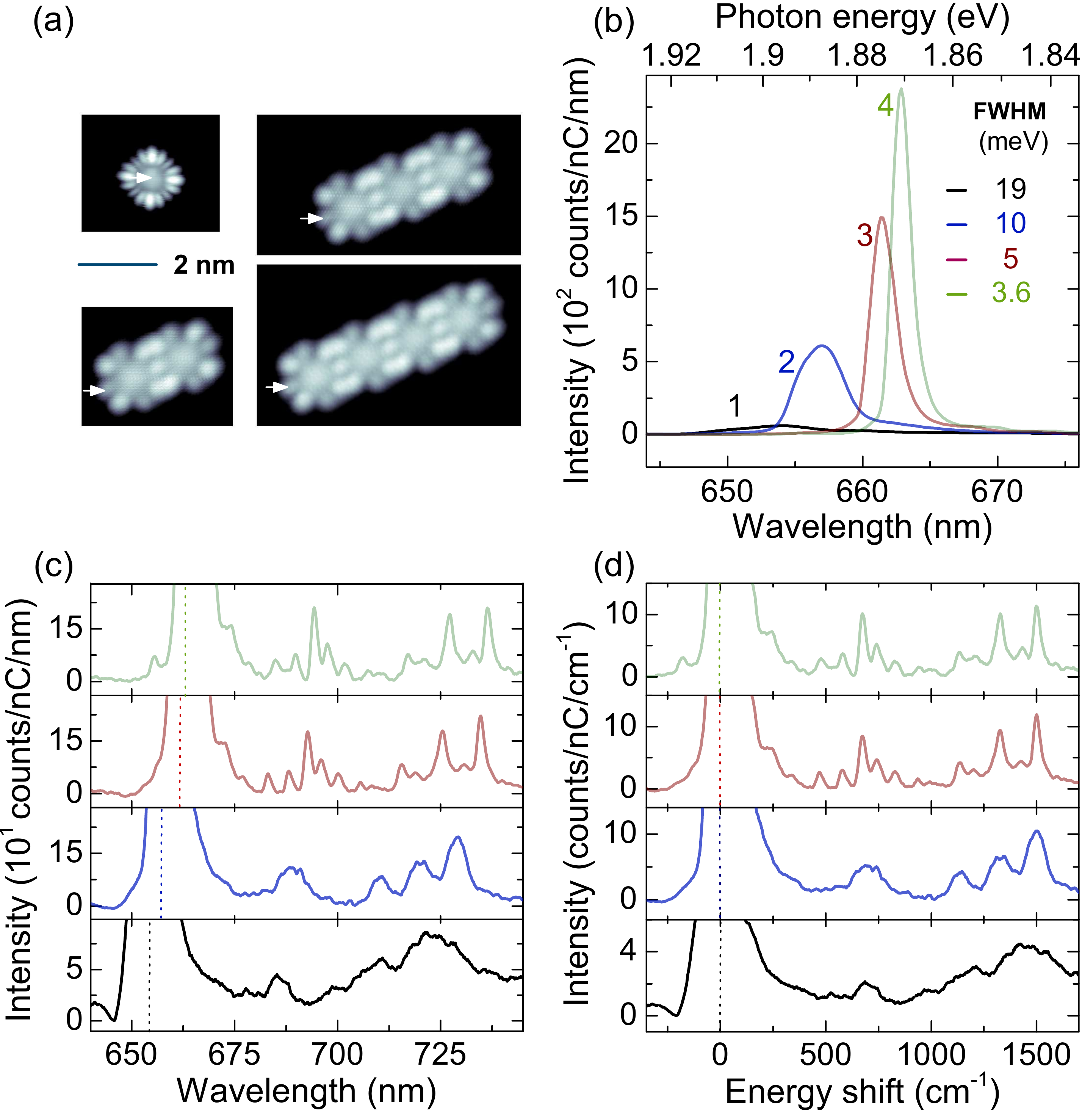}
  \caption{\label{fig2} (a) STM images ($I$ = 30 pA, $V$ = - 2.5 V) and (b) STM-LE spectra acquired for a single molecule, a dimer, a trimer and a tetramer of ZnPc molecules deposited on NaCl/Ag(111) ($V$ = -2.5 V). The white arrows in (a) mark the location of the tip during the acquisition of the optical spectra. The panels (c) and (d) show on the low energy vibrational features of the same spectra. The spectra are represented in (c) as a function of the wavelength and in (d) as a function of the shift from the 0-0 lines chosen as origin of the energies.} 
\end{figure}

In \Figref{fig2}(b) we display the STM-LE spectra recorded for linear arrangements made of 1 to 4 ZnPc molecules (\Figref{fig2}(a)). The 0-0 peak sharpens and shifts to lower energies with the number of molecules in the chain. These effects are the direct consequence of the coherent coupling of the molecular dipoles \cite{Zhang2016}. While the vibronic peaks appear at different wavelengths (\Figref{fig2}(c)) depending on the length of the linear arrangement, \Figref{fig2}(d) reveals an invariant energy shift of the vibronic modes with respect to the 0-0 line. Moreover, the widths of the vibronic peaks decrease when the number of molecules in the linear cluster increases, following the same trend observed for the 0-0 line. These observations suggest that the 0-0 and the vibronic emissions correspond to radiative transitions that all originate from the same excited state of the molecules.
                
\begin{figure}
  \includegraphics[width=1.00\linewidth]{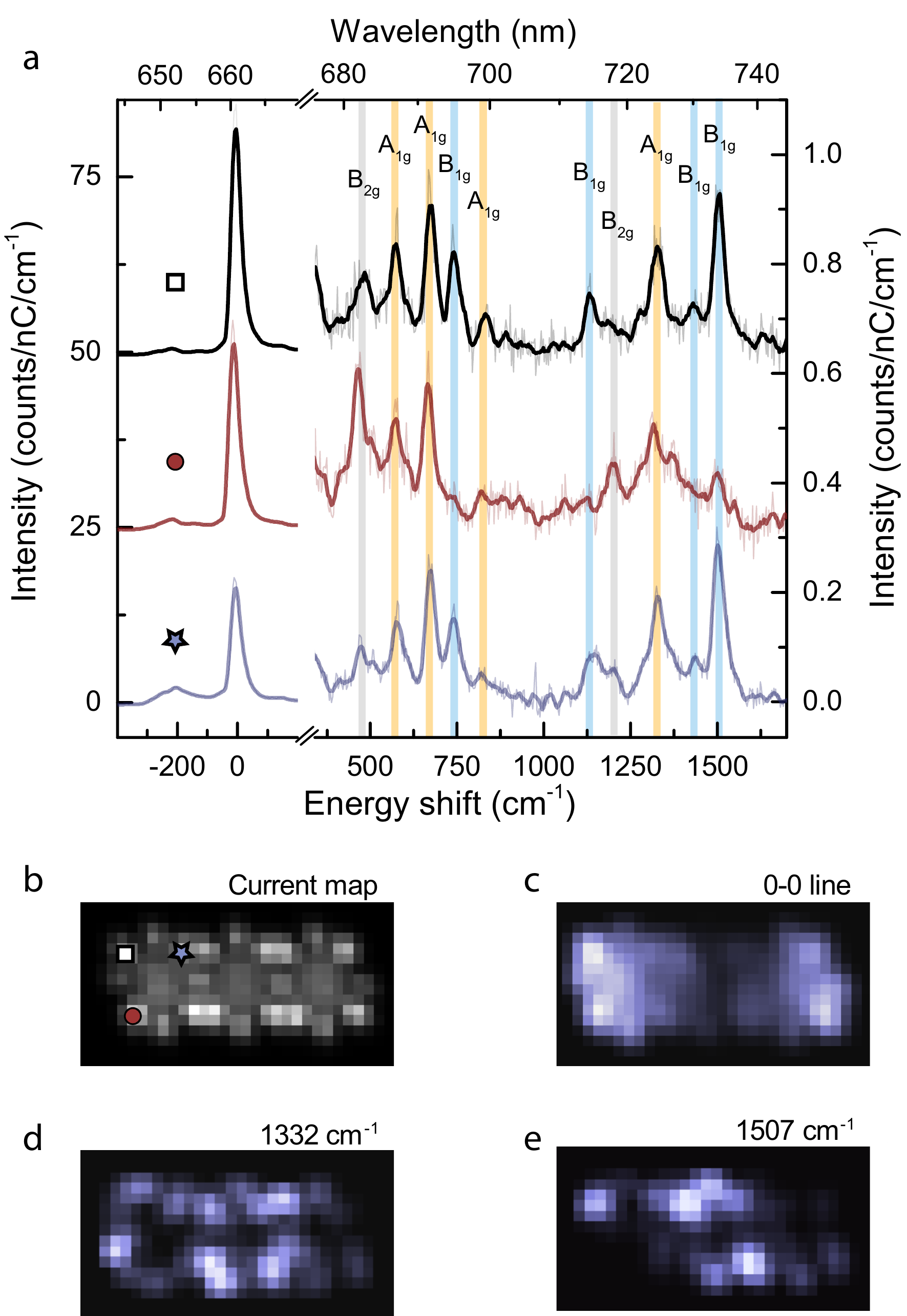}
  \caption{\label{fig3} (a) STM-induced light emission spectra ($V$ = - 2.5 V) acquired for three positions of the tip (coloured marks in (b)) with respect to a molecular trimer. The spectra are vertically shifted for clarity. (b) Constant height STM image of the trimer (5.1 $\times$ 2.7 nm$^2$, $V$ = - 2.5 V). Normalized photon intensity maps for (c) the 0-0 emission and for (d) the 1332 cm$^{-1}$ and (e) 1507 cm$^{-1}$ peaks (5.1 $\times$ 2.7 nm$^2$, $V$ = - 2.5 V, time/pixel = 10 s. The photon maps were slightly processed with a Gaussian smoothing algorithm.)} 
\end{figure} 
  
Having described how the emission is affected by the number of molecules in the assembly we will now focus on the spatial dependency of the vibronic signal. Figure \ref{fig3}(a) shows 3 spectra acquired for different locations of the STM-tip on a molecular trimer (square, dot, and star marks in \Figref{fig3}(b)). The intensity of the 0-0 line varies substantially as a function of the tip location. This effect has been described in details in Ref.\citenum{Zhang2016} and is another manifestation of the coupling between the molecular dipoles. The number and the energy of the vibronic peaks do not change, but the intensities of some peaks vary depending on the tip position. Here, three different behaviours can be distinguished: In a first group, the peaks (marked in yellow) show an almost constant intensity in the three spectra.  In a second group, the peaks (marked in blue) are intense in the top and bottom spectra and strongly attenuated in the middle spectrum. In a third group, the peaks (marked in grey) show the opposite pattern, with a low intensity in the top and bottom spectra and a high intensity in the middle spectrum. No obvious correlation with the intensity variation of the 0-0 line is observed, suggesting that the intensities of the vibronic peaks are weakly affected by the efficiency of the optical transition resulting in the 0-0 line. In the top panel, we indicate the irreducible representation of each mode as deduced by DFT calculations. This assignment is in agreement with earlier works \cite{Liu2007,Murray2011}. Interestingly, the three families of peaks are related to well-defined irreducible representations of the ZnPc molecule. The peaks having a constant intensity stem from ZnPc modes having A$_{1g}$ symmetry, while the second and third groups stem from ZnPc modes of A$_{2g}$ and B$_{1g}$ representations respectively. To get further insight into this phenomenon, we recorded maps of the light intensity at the energy of the 0-0 line (\Figref{fig3}(c)), for a vibronic peak whose intensity remains constant with tip position (h$\nu$ = 1332 cm$^{-1}$ ; A$_{1g}$ ; \Figref{fig3}(d)) and for a peak whose intensity varies (h$\nu$ = 1507 cm$^{-1}$ ; B$_{1g}$ ; \Figref{fig3}(e)). These maps were acquired with a constant tip-sample distance (open feedback loop) in order to prevent possible artefacts due to the tip trajectory. At each pixel of the map, an emission spectrum is recorded. A background is then subtracted to remove the contribution of the plasmonic emission. The spectra are eventually normalized by the tunneling current measured during the acquisition to correct for the higher (lower) excitation probability at high (low) current. The photon map at the energy of the 0-0 line (\Figref{fig3}(c)) shows two maxima at the extremities of the trimer, a behaviour that is attributed to the coherent coupling between the molecular dipoles \cite{Zhang2016}. This map is remarkably different from those acquired at the energy of the vibronic peaks (\Figref{fig3}(d,e)), confirming that the spatial dependencies of the vibronic peak intensities are not correlated to that of the 0-0 transition. Interestingly, the pattern in the vibronic photon maps is of higher symmetry for the mode at 1332 cm$^{-1}$ than for the mode at 1507 cm$^{-1}$, reflecting the higher symmetry of A$_{1g}$ compared to B$_{1g}$ modes.

In the following we provide an interpretation for the patterns in the photon maps. In \Figref{fig2}, a close relationship between the position and the width of the 0-0 line and the vibronic peaks is evidenced while \Figref{fig3} reveals an independent spatial variation of their intensities. These aspects may be reconciled assuming an interpretation of the spectral features in the frame of molecular vibronic coupling theory \cite{Fisher1984}. Indeed, in a purely Franck and Condon (FC) picture , one would expect the maps acquired at the energy of the vibronic peaks (Fig.3 d,e) to all faithfully reproduce the one of the 0-0 transition since all the lines in the spectrum stem from the same electronic transition and only their relative intensities are proportional to vibrational wavefunction overlap integrals (whose squares are the FC factors). A similar argumentation was developed \cite{Duan2015} to explain the strong resemblance between TERS maps acquired on a single molecule for different vibronic modes \cite{Zhang2013}. The spatial dependency of the vibronic peaks in \Figref{fig3} suggests that we need to move beyond the FC picture and to consider that the vibrational emission is provided by Herzberg-Teller contributions (\textit{i.e.}, the simultaneous change in electronic and vibrational states not due to simple vibrational overlaps). In this case the vibronic transitions directly depend on the contribution of electronic states of high energy that contaminate the lowest energy transition through the mechanism of vibronic coupling. This coupling directly depends on the vibrational mode considered, thus providing the relation between mode symmetry and the map patterns in \Figref{fig3}. A more detailed discussion of this mechanism is provided in supplemental materials \cite{SI}. 
\\

\begin{figure}
  \includegraphics[width=1.00\linewidth]{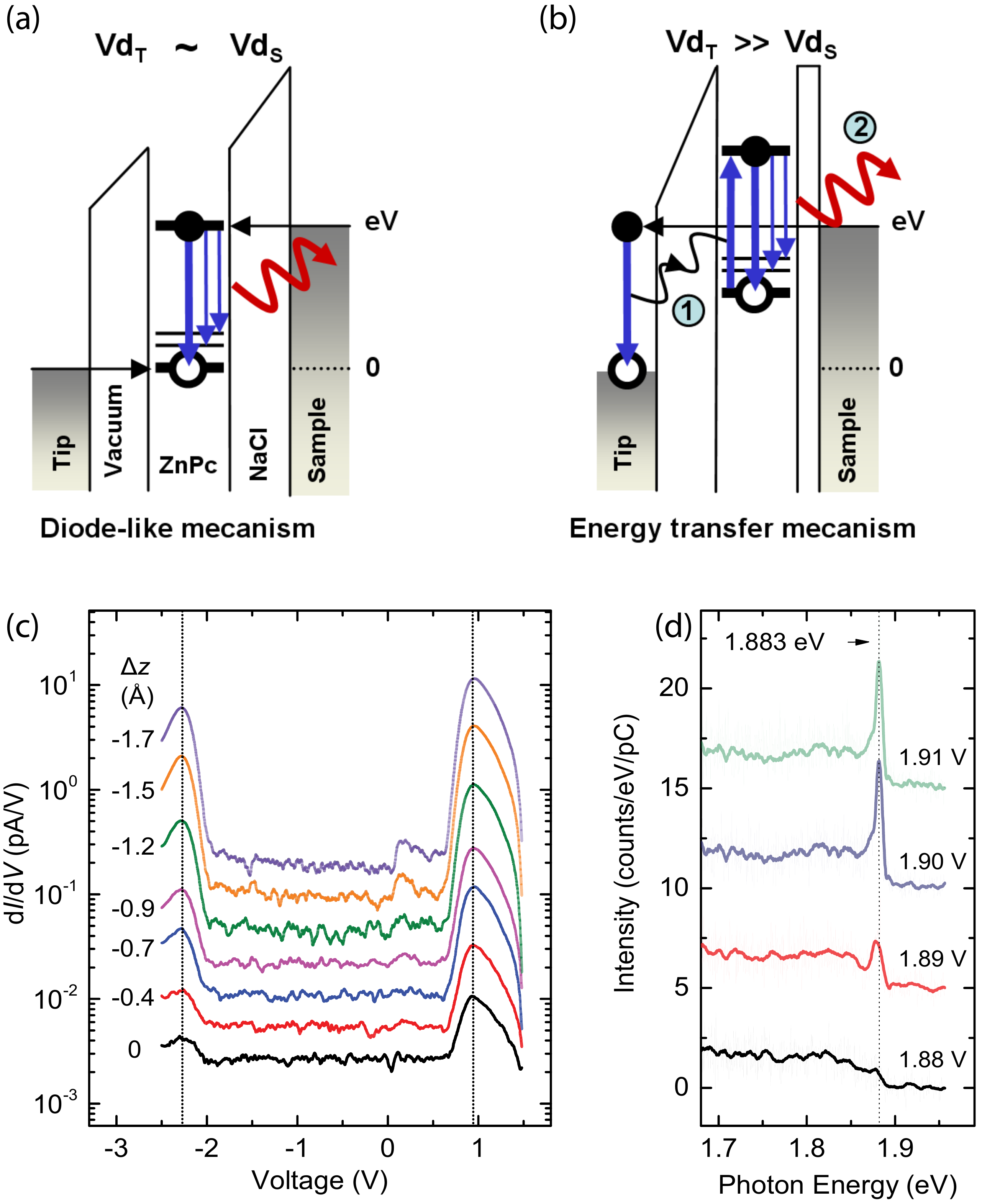}
 \caption{\label{fig4} Sketches (a) of the diode-like emission mechanism and (b) of the energy transfer mecanism. (c) Conductance d$I$/d$V$ spectra (in a logarithmic scale) acquired on a molecular dimer for different tip-molecule distances $\Delta$z. $\Delta$z = 0 corresponds to a current set point of $I$ = 7.5 pA at $V$ = - 2.5 V. (d) Light emission spectra acquired on the same dimer as a function of voltage ($I$ = 150 pA, $t$ = 180 s). The spectra are vertically shifted for clarity. } 
\end{figure} 

\noindent

Having formulated a tentative rationalization of the experimental observations, we will now try to critically discuss the excitation mechanism at play from an experimental point of view. STM-induced fluorescence of single-molecules has been reported in an increasing number of cases \cite{Qiu2003,Wu2008,Chen2010,Zhu2013,Lee2014,Reecht2014,Chong2016,Chong2016a,Zhang2016,Imada2016b,Imada2016}. In most of them (including Ref.\citenum{Zhang2016} on the same system) it was concluded that the luminescence involves a diode-like behaviour (\Figref{fig4}(a)) where electrons and holes injected in the molecule from the electrodes recombine and produce a photon. In this model, the frontier orbitals must shift with voltage with respect to both the Fermi levels of the tip and of the sample. Such a scenario happens when the voltage drop between the molecule and the surface (Vd$_\text{S}$) is of the same magnitude than between the molecule and the tip (Vd$_\text{T}$). However, conductance spectra acquired as a function of the tip-molecule distance show no shift of the frontier orbitals of the molecule (\Figref{fig4}(c)), a behaviour indicating that the voltage drops nearly exclusively on one side of the molecule (the tip side according to experiments performed on similar systems \cite{Repp2005}). This configuration is therefore not compatible with a diode-like mechanism. In contrast, the spectra in \Figref{fig4}(d) show that the voltage onset of the 0-0 line exactly matches the energy of the 0-0 transition. This rather suggests an energy transfer (\Figref{fig4}(b)) between the tunneling electrons and the molecule \raisebox{.5pt}{\textcircled{\raisebox{-.9pt}{1}}} mediated by the plasmons localized at the tip-sample junction. The excited molecule then relaxes to the ground or to an excited vibrational state of the electronic ground state by emitting a photon \raisebox{.5pt}{\textcircled{\raisebox{-.9pt}{2}}}. A similar interpretation was first proposed to explain the luminescence spectra of large molecular aggregates \cite{Dong2010,Schneider2012a} and the emission of single molecules suspended in a STM junction \cite{Chong2016,Chong2016a}. This mechanism was nicely confirmed for single molecules adsorbed flat on an insulating layer \cite{Imada2016}.\\      
In conclusion, our experiment provides a unique way to obtain a high chemical sensitivity with sub-molecular scale spatial resolution without resorting to an optical excitation. By using electrons rather than photons to excite the low-temperature fluorescence of ZnPc molecules, vibronic signals with sub-molecular resolution were obtained. This approach should allow distinguishing different chromophores co-adsorbed on a same surface. In contrast to resonant TERS experimental maps \cite{Zhang2013} that revealed to be insensitive to the vibronic modes \cite{Duan2015}, the patterns of our STM-LE ''vibronic maps'' is mode-dependent and highlight the importance of its symmetry. This behaviour can be rationalized by considering the different origin of the vibronic emission in these two experiments. While the resonant TERS signals finds its origin in the dominant character of the Franck-Condon term, Herzberg-Teller contributions prevail in our spectra. Our approach therefore constitutes a decisive step towards a vibronic mode spectroscopy of molecules with sub-nanometric resolution.\\ 

\noindent
The authors thank Jean-Louis Gallani for the absorption and fluorescence spectra of the ZnPc molecules in solution and Virginie Speisser, Jean-Georges Faullumel and Olivier Cregut for technical support. The Agence National de la Recherche (project SMALL'LED No. ANR-14-CE26-0016-01), the Labex NIE (Contract No. ANR-11-LABX-0058\_NIE), the R�gion Alsace and the International Center for Frontier Research in Chemistry (FRC) are acknowledged for financial support.

\end{document}